\documentclass[runningheads]{llncs}
\usepackage{algorithm}
\usepackage{algpseudocode}
\usepackage{amsmath}
\usepackage{booktabs}
\usepackage{color}
\usepackage[T1]{fontenc}
\usepackage{graphicx}
\usepackage[bookmarksdepth=2]{hyperref}
\usepackage[utf8]{inputenc}
\usepackage{subcaption}
\usepackage{comment}
\usepackage{tikz}
\usepackage{xspace}
\usepackage{color}
\usepackage{amssymb}
\usepackage{listings}
\usepackage{cite}

\usepackage{version}

\usepackage[firstpage]{draftwatermark}

\SetWatermarkAngle{0}
\SetWatermarkText{\raisebox{12.7cm}{
\hspace{-0.30cm}
\href{https://doi.org/10.1109/5.771073}{\includegraphics{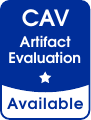}}
\hspace{8cm}
\includegraphics{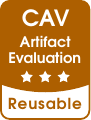}
}}

\includeversion{conf}
\excludeversion{rep}

\let\llncssubparagraph\subparagraph
\let\subparagraph\paragraph
\usepackage[compact]{titlesec}
\let\subparagraph\llncssubparagraph
\titlespacing*{\section}{0pt}{*2.0}{*2.0}
\titlespacing*{\subsection}{0pt}{*1.5}{*1.0}
\titlespacing*{\subsubsection}{0pt}{*0.6}{*0.4}

\definecolor{keywordcolor}{rgb}{0.7, 0.1, 0.1}
\definecolor{tacticcolor}{rgb}{0.0, 0.1, 0.6}
\definecolor{commentcolor}{rgb}{0.4, 0.4, 0.4}
\definecolor{symbolcolor}{rgb}{0.0, 0.1, 0.6}
\definecolor{sortcolor}{rgb}{0.1, 0.5, 0.1}
\definecolor{attributecolor}{rgb}{0.7, 0.1, 0.1}

\DeclareUnicodeCharacter{03B1}{$\alpha$}
\DeclareUnicodeCharacter{2200}{$\mathbb{\forall}$}
\DeclareUnicodeCharacter{2264}{$\mathbb{\le}$}
\DeclareUnicodeCharacter{2081}{$_{1}$}
\DeclareUnicodeCharacter{2082}{$_{2}$}

\def\orcidID#1{\smash{\href{http://orcid.org/#1}{\protect\raisebox{-1.25pt}{\protect\includegraphics{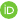}}}}}

\hypersetup{
 pdfborder={0 0 0},
 colorlinks=true,
 linkcolor=gray,
 urlcolor=gray,
 citecolor=gray,
}

\newcommand{\cvcv}{\textsc{cvc5}\xspace}
\newcommand{\tactic}{\textsc{lean-smt}\xspace}
\newcommand{\auto}{\textsc{lean-auto}\xspace}
\newcommand\HOLyHammer{HOL\kern-.075ex\raise.19ex\hbox{\textsc{y}}\kern-.05ex\-Hammer}

\def\mytitle{\tactic: An SMT tactic for discharging proof goals in Lean}

\title{\mytitle}
\author{
Abdalrhman Mohamed\inst{1}\orcidID{0000-0003-1414-7073}
\and Tomaz Mascarenhas\inst{4}\orcidID{0000-0003-2747-8349}
\and Harun Khan\inst{1}\orcidID{0000-0003-3379-5631}
\and Haniel Barbosa\inst{4}\orcidID{0000-0003-0188-2300}
\and Andrew Reynolds\inst{2,3}\orcidID{0000-0002-3529-8682}
\and Yicheng Qian\inst{1}\orcidID{0009-0008-0194-9572}
\and \\ Cesare Tinelli\inst{2}\orcidID{0000-0002-6726-775X}
\and Clark Barrett\inst{1}\orcidID{0000-0002-9522-3084}
}
\titlerunning{Lean-SMT: An SMT tactic for discharging proof goals in Lean}
\authorrunning{Mohamed et al.}
\institute{
  Stanford University, Stanford, USA
  \and
  The University of Iowa, Iowa City, USA
  \and
  Amazon Web Services, Seattle, WA, USA
  \and
  Universidade Federal de Minas Gerais, Belo Horizonte, Brazil
}

\begin{document}

\maketitle

\begin{abstract}
  Lean is an increasingly popular proof assistant based on dependent type
  theory. Despite its success, it still lacks important automation features
  present in more seasoned proof assistants, such as the Sledgehammer tactic in
  Isabelle/HOL.
  A key aspect of Sledgehammer is the use of proof-producing SMT solvers to prove
  a translated proof goal and the reconstruction of the resulting proof into valid
  justifications for the original goal.
  We present \tactic, a tactic providing this functionality in Lean.
  We detail how the tactic converts Lean goals into SMT problems and, more
  importantly, how it reconstructs SMT proofs into native Lean proofs.
  We evaluate the tactic on established benchmarks used to evaluate
  Sledgehammer's SMT integration, with promising results.
  We also evaluate \tactic as a standalone proof checker for proofs of SMT-LIB
  problems.  We show that \tactic offers a smaller trusted core without
  sacrificing too much performance.
\end{abstract}

\section{Introduction}

Proof assistants, also known as interactive theorem provers (ITPs), allow users
to write mechanized proofs of statements written in a formal language,
whose validity can be verified by a small, trusted kernel.
They help users construct trustworthy, formal, machine-checkable proofs of theorems, and have
been increasingly used to mechanize proofs of various mathematical results~\cite{fourColors,kepler}.
This process has significantly accelerated in recent years with the
adoption of the Lean proof assistant~\cite{deMoura2021} by leading members of
the mathematical community~\cite{scholze,mathlib,Tao2025}.
Proof assistants are also commonly used in certain areas of computer science to
model and formally verify systems, thanks to the high
expressiveness of their underlying language and logic~\cite{Klein2010,Nelson2020}.

The trustworthiness of proof assistants relies on the kernel correctly verifying
every proof step.  For this reason, ITP kernels are designed to be simple and
small, implementing just
straightforward logical operations from the logical framework underlying the proof
assistant.
This means that, in principle, each proof step must be explicitly formulated by
the user, so that naive use of ITPs would require a tremendous amount of
expertise and effort.
A major challenge, then, is the extension of the kernel with trustworthy
facilities for automating the
writing of mechanized proofs as much as possible, thereby reducing the burden on the users.%
\footnote{In mathematics, the cost of mechanizing an informal
  proof is currently estimated to be $\sim 20$x~\cite{Tao2025}.
  Using ITPs to formally verify large systems is well known to be very costly,
  in the range of multiple person-years~\cite{Klein2010}.  }
Automation is generally achieved via \emph{tactics}, proof-producing
algorithms, traditionally written in a special-purpose language, that
discharge proof obligations for certain classes of subgoals, often by
simulating common proof techniques, such as case analysis or induction.

An alternative is
to use external automated theorem provers (ATPs) to solve subgoals
when possible.
Tools such as \HOLyHammer~\cite{Kaliszy2015},
Miz$\mathbb{AR}$~\cite{Jakubuv2023},
Sledgehammer~\cite{DBLP:journals/iandc/MengQP06}, and Why3~\cite{Bobot2011},
provide a one-click connection from proof assistants to first-order provers and
have led to considerable improvements in proof-assistant
automation \cite{Blanchette2016-qed}.
The Sledgehammer tactic in Isabelle/HOL~\cite{Nipkow2002} is particularly
successful and includes an integration with
proof-producing SMT
solvers~\cite{DBLP:conf/cade/BlanchetteBP11,DBLP:conf/cade/SchurrFD21}.  Sledgehammer
translates certain proof goals into SMT-LIB~\cite{Barrett2017}, a standard
input format for SMT solvers, and sends them to the solver.  The resulting proof
is reconstructed by essentially reproducing each step within the proof assistant.
%
This extension
to Sledgehammer has been especially useful for proof goals arising from
formal verification efforts carried out in
Isabelle/HOL~\cite{DBLP:conf/cade/BlanchetteBP11}.
We conjecture that any tactic in Lean with the same ambition and potential impact
as Sledgehammer will require a similar level of SMT solver integration.

In this paper, we present \tactic,\footnote{\tactic is available online
  at \url{https://github.com/ufmg-smite/lean-smt}} which provides an initial
implementation of a similar integration in Lean and is a stepping stone towards a full
Sledgehammer-like tactic for Lean.
\tactic operates by translating Lean proof goals expressible in the first-order
logic fragment of dependent type theory (Section~\ref{sec:translation}) into SMT problems,
leveraging SMT theories to model corresponding elements from Lean, such as
uninterpreted functions, propositional equality, first-order quantifiers,
and arithmetic operators.
To bridge the gap between this restricted fragment and the goals that arise in
practice in Lean formalizations, we rely on the
\auto tactic\footnote{\url{https://github.com/leanprover-community/lean-auto}}
to reduce proof goals to first-order logic, together with dedicated preprocessing in
\tactic itself to express them in the language of selected SMT theories (Section~\ref{sec:preprocessing}).
\tactic currently supports the state-of-the-art proof-producing SMT solver
cvc5~\cite{Barbosa2022-cvc5}.
cvc5's extensive proof production capabilities~\cite{Barbosa2022} and
strong performance on the fragment of interest make it well suited
for such an integration.
cvc5's proofs are reconstructed into native Lean proofs
(Section~\ref{sec:proofRec}) by using either a
Lean theorem, a Lean tactic, or a formally verified Lean program.

We evaluate \tactic (Section~\ref{sec:eval}) on proof goals from a standard
benchmark set used to evaluate Sledgehammer, showing that \tactic performs
comparably with Sledgehammer's SMT integration.
We also evaluate it as a verified proof checker for SMT proofs. While, as expected, \tactic is less performant than standalone, unverified SMT proof checkers, its performance is generally within an order of magnitude of theirs. Additionally, it offers comparable performance to SMTCoq~\cite{smtcoq}, a similarly verified checker for Coq, while supporting a larger logical fragment.

\section{Related Work}

While our main inspiration for \tactic has been the SMT integration in
Sledgehammer, there are other ways in which Sledgehammer leverages the power of
automatic theorem provers.
Sledgehammer includes a premise selection module, which filters lemmas from
Isabelle's libraries that are potentially useful for proving the goal. These
lemmas are added to the problem to be sent to the solver.
Once the problem is solved, rather than directly reconstruct the proof, an
alternative strategy is to use it only to identify the lemmas used to prove the goal
and pass these lemmas, together with the original goal, to
\emph{metis}~\cite{metis}, a proof-producing superposition prover written as an
Isabelle tactic.
While there is no equivalent for Sledgehammer in Lean, there is a growing
ecosystem of tools that eventually can be combined into a comparable tool.
Recently Aesop~\cite{aesop}, a tableaux-based prover, and Duper~\cite{duper}
which is, like \emph{metis}, a superposition prover, were introduced as
proof-producing Lean tactics.
No equivalent for the premise selection mechanism is readily available in Lean yet,
although there has been initial work in this direction~\cite{premiseSelection}.

Another integration between a proof assistant and SMT solvers is
SMTCoq~\cite{smtcoq}, within the Coq proof assistant~\cite{coq}.
It supports proof reconstruction for the SMT solvers veriT~\cite{verit} and
CVC4~\cite{cvc4}, but rather than replaying each individual proof step within
Coq, as Sledgehammer and \tactic do, it applies a formally verified checker
that, if successful, confirms the original proof goal as a theorem.
This approach relies heavily on the efficiency of the proof assistant itself,
since the proof checker runs within the proof assistant and
may have to analyze and simplify huge proof terms. The Coq proof assistant is designed to
be very fast at this, while Lean, on the other hand, is not~\cite{baanenPhd}.
This observation contributed to the decision to use proof replay in
\tactic.
Moreover, the verified checker approach is rather rigid, since any modification
to the supported proof format requires the checker's correctness theorem to
be proven again, which can be a significant effort.
In the proof replay approach, one needs only to change the reconstruction tactic
by modifying the step corresponding to the changed element of the format.
This has been especially useful while developing \tactic since cvc5's proof
calculus and infrastructure are still evolving
~\cite{Barbosa2022,Notzli2022,Lachnitt2024}.

\section{System Overview}

\begin{figure}[t]
    \centering
    \includegraphics[width=0.5\textwidth]{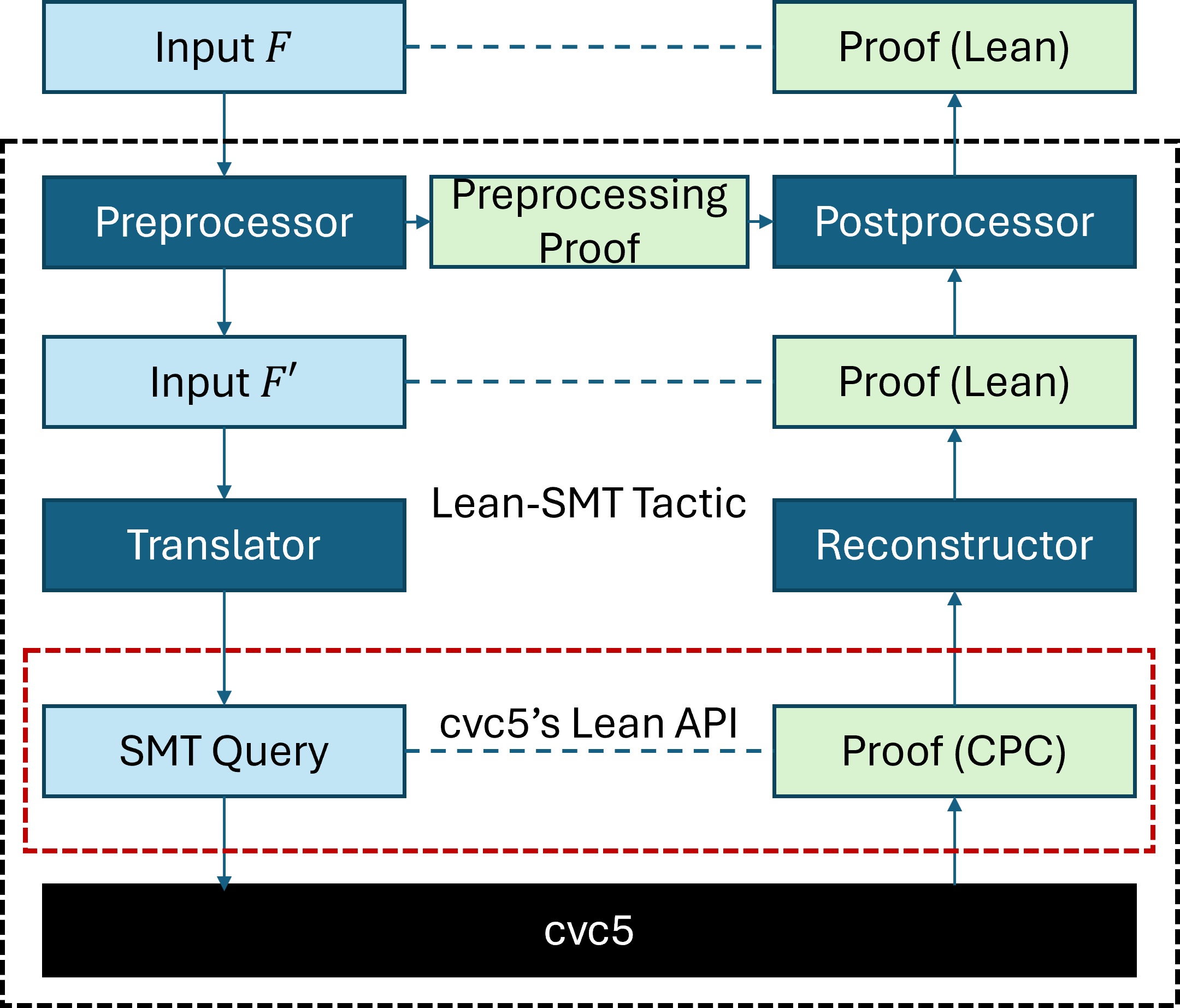}
    \caption{Architecture of the \tactic tactic.}
    \label{fig:arch}
\end{figure}

Figure~\ref{fig:arch} depicts the \tactic architecture, which takes as
input a Lean goal, whose type is represented as the formula $F$, and
generates a proof for it by solving a corresponding SMT problem and
reconstructing the proof into a Lean proof for $F$.
The arrows illustrate the tactic's pipeline, through which the input formula is
simplified, translated into an SMT query, and sent to cvc5.  The solver's proof
is then reconstructed as a proof in Lean for the input formula.
The light-green boxes represent the proofs used during the reconstruction
stage, which correspond directly to the formulas (light-blue) processed during
the translation stage. The dashed lines emphasize this correspondence.

The \emph{preprocessor} module converts the initial input \( F \) into an
intermediate form \( F' \), simplifying or restructuring the input to make it
more suitable for translation into an SMT query. During this phase, a
preprocessing proof is generated, capturing the transformations applied.
The \emph{translator} module generates a formula in SMT-LIB whose unsatisfiability corresponds to the
validity of $F'$.
The formula is passed to a solver (cvc5) that evaluates it and produces a proof
if the formula is determined to be unsatisfiable, outputting a proof in the
Cooperating Proof Calculus (CPC)
format.\footnote{See
  \url{https://cvc5.github.io/docs/cvc5-1.2.1/proofs/output_cpc.html}.
  A complete list of the proof rules in CPC can be found
  at
  \url{https://cvc5.github.io/docs/cvc5-1.2.1/api/cpp/enums/proofrule.html}.
  The semantics of the rules is also defined in the Eunoia logical framework,
  described in the user manual of the Ethos proof
  checker: \url{https://github.com/cvc5/ethos/blob/main/user\_manual.md}.}
The tactic interfaces with cvc5 through Lean's Foreign Function Interface (FFI)
and the solver's Lean API, which we added to the solver to facilitate the
integration.
The \emph{reconstructor} module translates the CPC proof into a Lean proof of
$F'$, mapping the CPC proof structure to corresponding Lean constructs and
ensuring logical equivalence.
The \emph{postprocessor} combines the preprocessing proof and the
reconstruction proof into a single Lean proof for the original formula $F$,
which is then checked by Lean's kernel.

\subsection{Preprocessing Original Lean Goal}
\label{sec:preprocessing}

Lean's type system, rooted in dependent type theory (DTT)~\cite{theoremprovinginlean},
is far more expressive than the many-sorted first-order logic
(FOL)~\cite{Enderton2001} used by SMT solvers.
To bridge this gap, we employ proof-producing preprocessing steps
that simplify Lean goals into a form more amenable to translation into SMT-LIB.

We first use
\auto,\footnote{\url{https://github.com/leanprover-community/lean-auto}}
to reduce the goal to FOL.
\auto normalizes universe levels,
monomorphizes definitions, adds lemmas related to inductive types, and
replaces type class instances by their corresponding values.
All of these transformations are implemented in Lean and are proof producing, so their soundness is guaranteed
by Lean's kernel.
However, they are inherently incomplete, given the expressivity gap between DTT
and FOL.

The preprocessing applied by \auto is not specific to SMT solvers, so while the
resulting goal is in FOL, it is not aligned with the SMT-LIB standard.
We apply a further preprocessing step so that certain Lean types and constructs
(e.g., \texttt{Prop}, \texttt{Nat}, \texttt{Rat}, and \texttt{Iff}) that don't have
direct counterparts in SMT-LIB can be transformed into types and constructs
that do.

\begin{example}
Consider the following Lean goal, which asserts the uniqueness of the identity element in a group:
\begin{lstlisting}
⊢ ∀ (G : Type u) [Group G] (e : G), (∀ (a : G), e * a = a) ↔ e = 1
\end{lstlisting}
This goal cannot be directly translated into SMT-LIB due to the presence of the type class \texttt{Group} and the logical operator \( \leftrightarrow \).
During preprocessing, the goal is transformed and expanded to:
\begin{lstlisting}
G: Type u
inst: Group G
e e': G
op: G → G → G
inv: G → G
one_mul: ∀ (a : G), op e a = a
inv_mul_cancel: ∀ (a : G), op (inv a) a = e
mul_assoc: ∀ (a b c : G),
op (op a b) c = op a (op b c)
⊢ (∀ (a : G), (op e' a = a)) = (e' = e)
\end{lstlisting}
Here, \auto replaces the type class \texttt{Group} with explicit assumptions about the group operations (e.g., associativity, identity, and inverse axioms), and \tactic's preprocessing transforms the bidirectional logical operator \( \leftrightarrow \) into a suitable equality comparison. These transformations make the goal compatible with SMT-LIB's logic while preserving its original meaning.
\end{example}

\subsection{Translation to SMT-LIB}
\label{sec:translation}

After preprocessing, translating Lean goals into SMT-LIB is relatively straightforward, as the fragments mostly overlap. However, one key challenge stems from differing assumptions about sorts: SMT-LIB assumes sorts are non-empty, while Lean allows types to be empty.
This discrepancy can make the translation unsound.
The reconstruction stage ensures
soundness by failing if a proof step depends on a type being non-empty and Lean
cannot establish that the type is an instance of the type class of non-empty
types.
As long as the instances are found, this discrepancy between the logic systems
is successfully addressed.

\begin{example}
Continuing from our previous example, the preprocessed goal is translated into SMT-LIB as follows:
\begin{lstlisting}[language=LISP]
(declare-sort G 0)
(declare-const e G)
(declare-const |e'| G)
(declare-fun op (G G) G)
(declare-fun inv (G) G)
(assert (forall ((a G)) (= (op e a) a)))
(assert (forall ((a G)) (= (op (inv a) a) e)))
(assert (forall ((a G) (b G) (c G)) (= (op (op a b) c) (op a (op b c)))))
(assert (distinct (forall ((a G)) (= (op |e'| a) a) (= |e'| e))))
(check-sat)
\end{lstlisting}
Note that the translation is sound in this example because any group \( G \) is guaranteed to be non-empty due to the existence of the identity element \( e \).
\end{example}

\subsection{cvc5's Proof Format and Reconstruction}
\label{sec:proofRec}

When cvc5 establishes that an SMT query is unsatisfiable, it can optionally
generate a proof in the CPC format that accurately mirrors its internal
reasoning processes.
In CPC, each proof rule can be represented as follows:

\[
\frac{\varphi_1 \cdots \varphi_m \mid t_1 \cdots t_n}{\psi}{C},
\]
\noindent
where $\varphi_1, \ldots, \varphi_m$ are the premises, $t_1, \ldots, t_n$ are
the arguments provided to the proof rule, and $C$ (which may be absent) denotes a decidable side-condition.
The proof format currently specifies over 662 proof rules in various domains,
including arithmetic, strings, quantifiers, and higher-order logic. \tactic supports
around 200 of these proof rules, which amounts to approximately 30\%. We prioritized this subset because it suffices to support the most common proof goals in Lean (it corresponds to the same logical fragment supported initially by Sledgehammer and SMTCoq), with the other rules corresponding to the reasoning of specific theories.
To reconstruct the cvc5 proofs in Lean, \tactic processes the CPC proof step by step,
translating each proof step into an equivalent one in Lean.
After \tactic completes proof reconstruction, the entire proof is submitted to
the Lean kernel for verification.
In cases where a proof step cannot be reconstructed, it is presented to the user
to be proved manually, ensuring that the reconstruction process remains sound.
Also note that since the logic in SMT-LIB is classical, certain parts of the
proof rely heavily on the axiom of choice.
Below we detail each of the reconstruction techniques we apply.

\paragraph{Reconstruction via theorems.} Proof rules without side conditions
generally correspond directly to theorems in Lean. We
proved an extensive library of such theorems, which cover 163 proof rules, to use for proof reconstruction.

\begin{example}
Consider the \texttt{ARITH\_MULT\_TANGENT} proof rule from CPC.
\begin{align*}
    \frac{- \mid x, y, a, b, \sigma}{(t \leq tplane) = ((x \le a \wedge y \ge b) \vee (x \ge a \wedge y \le b))} \text{if } \sigma = \top\\
    \frac{- \mid x, y, a, b, \sigma}{(t \leq tplane) = ((x \le a \wedge y \le b) \vee (x \ge a \wedge y \ge b))} \text{if } \sigma = \bot
\end{align*}
where $x,y$ are real terms, $t=x\cdot y, a, b$ are real constants, $\sigma \in \{ \top, \bot\}$ and $tplane := b \cdot x + a\cdot y-a\cdot b$ is the tangent plane of $x\cdot y$ at $(a, b)$.
Formalizing this proof rule into a Lean theorem is straightforward:
\begin{lstlisting}
theorem arithMulTangentLowerEq :
(x * y ≤ b * x + a * y - a * b) = ((x ≤ a ∧ y ≥ b) ∨ (x ≥ a ∧ y ≤ b))
theorem arithMulTangentUpperEq :
(x * y ≤ b * x + a * y - a * b) = ((x ≤ a ∧ y ≤ b) ∨ (x ≥ a ∧ y ≥ b))
\end{lstlisting}

\end{example}

The formalization of other proof rules is more complex and often requires design
choices when stating the theorem to capture the semantics of the corresponding
CPC rule.
\begin{example}
Consider for example the \texttt{RESOLUTION} proof rule:
\begin{align*}
    \frac{C_1 \quad C_2 \mid pol, L}{C},
\end{align*}
where $C_1, C_2$ are disjunctions, $L$ is a disjunct occurring positively
(respectively, negatively) in $C_1$ and negatively (resp. positively) in $C_2$
if $pol$ is the Boolean \emph{true} (resp. \emph{false}).
The result $C$ is a disjunction with the disjuncts from $C_1$ minus $L$
(resp. $\neg L$) and the disjuncts from $C_2$ minus $\neg L$ (resp. $L$) if
$pol$ is the Boolean \emph{true} (resp. \emph{false}).
Moreover $C$ is such that the remaining disjunctions from $C_1$ and $C_2$ occur
as a chain of disjunctions between each disjunct as opposed to nested disjunctions.
These restrictions capture the view, within cvc5, of disjunction as an
``$n$-ary'' operator, which takes multiple arguments.
To capture this reasoning in Lean, where disjunction is a binary operator, one
needs to carefully reason about associativity and commutativity to exactly
reproduce the semantics of this rule. We encapsulate it in a Lean theorem as follows:
\begin{lstlisting}
theorem orN_resolution (hps : orN ps) (hqs : orN qs)
    (hi : i < ps.length) (hj : j < qs.length)
    (hij : ps[i] = ¬qs[j]) :
    orN (ps.eraseIdx i ++ qs.eraseIdx j)
\end{lstlisting}

The premises, \lstinline{hps : orN ps} and \lstinline{hqs : orN qs}, are
disjunctions, but built with the \lstinline{orN} function which takes
a \lstinline{List} of literals represented as \lstinline{Prop}.
This formulation avoids representing \lstinline{ps} and \lstinline{qs} as for
example inductively defined \lstinline{Prop} instead of \lstinline{List} or
encoding the literals as \lstinline{Bool} instead of \lstinline{Prop}.
Using \lstinline{List} permits leveraging general theorems
about \lstinline{List} we proved for eliminating tedious corner cases that would
otherwise arise in a \lstinline{Prop} implementation.
Additionally, the choice to encode Boolean literals as \lstinline{Prop} is a
deliberate choice due to the type-theoretic foundation of Lean.
\end{example}

\paragraph{Reconstruction via tactics.} Some proof rules without side
conditions require the application of multiple lemmas or more complex reasoning.
We implement specialized tactics to encapsulate these steps. This streamlines
the reconstruction process by automating repetitive or intricate reasoning
steps. We cover 37 proof rules this way, which required implementing a
library with around 400 theorems.

\begin{example}
  A proof rule that is very general, making it hard to state and prove as a
  theorem, is the \texttt{ARITH\_SUM\_UB} rule:
  \begin{align*}
    \frac{\bigwedge_{i = 1}^{n} a_{i} \bowtie_{i} b_{i} \mid a_i, b_i}{\sum_{i = 1}^{n} a_{i} \bowtie^{*} \sum_{i = 1}^{n} b_{i}}
  \end{align*}
  where $\bowtie_i$ can be either $<$, $\leq$ or $=$, and $\bowtie^{*}$ is
  either $<$, if at least one of the $\bowtie_i$ is $<$, or $\leq$, otherwise.
  Moreover, while each pair of variables
  $a_i$ and $b_i$ always have the same type, it is possible that different
  pairs have different types, some being integers and some being reals.
  It is possible to encode this proof rule as a single theorem in Lean, but the
  statement of the theorem would be quite intricate, due to the necessity of
  lifting the integer variables to reals and of combining the inequalities
  statically.
  Also, it is likely that this would make it very hard to prove.
  In this case, it is easier to write a tactic that considers the different
  cases of the rule and applies an appropriate, simpler theorem for each case.
  The implementation of this tactic requires 9 variations of the following theorem:

\begin{lstlisting}
  sumBoundsThm {α : Type} [LinearOrderedRing α] {a b c d : α} :
    a < b → c < d → a + c < b + d
\end{lstlisting}

  Each variation corresponds to one possible combination of the inequality symbols in the hypothesis. The relation symbol in the conclusion is
  adapted accordingly in each theorem. Since the rule accepts mixing of variables
  from Int and Real types, we need a variation of each one of those 9 theorems for each combination
  of the types of the variables. Instead of stating all the combinations explicitly, which would
  result in a total of 36 theorems and a long branch in the implementation of the tactic,
  we state only one polymorphic version of each, as indicated by the type parameter \texttt{α} in \texttt{sumBoundsThm}.
  Obviously, these theorems do not hold
  for any \texttt{α} (it cannot even be stated if there are no comparison or addition operators defined over \texttt{α}).
  We solve this issue by adding a restriction, stating that \texttt{α} satisfies the axioms of a
  \textit{Linear Ordered Ring} (a class of types that contains both \texttt{Int} and \texttt{Real} defined in mathlib).
  With this restriction each theorem can be proven. The full tactic can be
  seen in Appendix~\ref{sec:rec-tactic}.

\end{example}

\paragraph{Reconstruction via reflection.} For proof rules involving complex side conditions or computations (such as
arithmetic simplifications), we encode the side condition into a reflective
decision procedure, which is formally verified in Lean. The proof rule is then translated into a theorem with
the side condition as an additional premise. Applications of this rule are
verified by the Lean kernel using definitional equality.
We cover 5 proof rules this way. We reused one program from Lean's library,
\texttt{ac\_rfl}, which applies associative and commutative properties of
addition and multiplication to normalize arithmetic expressions; and we implemented
a new program, \texttt{poly\_norm}, which normalizes polynomials up to
associativity, commutativity, and distributivity by expanding polynomials.

\begin{example}

Consider the example:
\begin{lstlisting}
example (x y : Int) (z : Real) :
  1 * ↑(x + y) * z / 4 = 1 / (2 * 2) * (z * ↑y + ↑x * z) := by poly_norm
\end{lstlisting}
\end{example}

Proving the correctness of \texttt{poly\_norm} required proving around 70
theorems, amounting to 620 lines of Lean code.
We define a monomial as an ordered list of natural numbers representing variable indices, so that equality of two monomials is immediate.
Polynomials are defined as lists of monomials, and theorems about monomials generalize to polynomials using induction.
We define \lstinline{denote}, which essentially evaluates polynomial expressions.
Using lemmas we proved about lists and the objects we defined, we prove a theorem pushing \lstinline{denote} into each operator.
\begin{lstlisting}
theorem denote_mul {p q : Polynomial} :
  (p.mul q).denote ctx = p.denote ctx * q.denote ctx
\end{lstlisting}
Proving similar theorems for each operator (addition, multiplication, division by a constant, and negation) yields a correctness theorem.
\begin{lstlisting}
theorem denote_eq_from_toPoly_eq {e₁ e₂ : RealExpr}
  e₁.toPoly = e₂.toPoly → e₁.denote ictx rctx = e₂.denote ictx rctx
\end{lstlisting}
Since variables are represented as natural numbers, the premise of the theorem does not contain actual variables. Therefore, Lean can establish the premise through definitional equality. Moreover, the premise is decidable, allowing us to compile it into machine code for enhanced performance. In our experiments, this approach achieved speeds up to 25 times faster than using definitional equality for very large arithmetic expressions.

\section{Evaluation and Results}
\label{sec:eval}

\begin{figure}[t]
  \centering
  \begin{subfigure}[b]{0.49\textwidth}
  \centering
  \includegraphics[width=\textwidth]{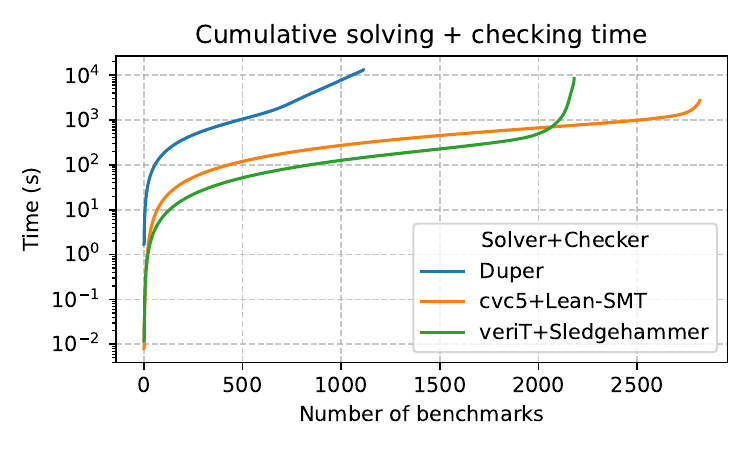}
  \caption{}
  \label{fig:sledgehammer}
  \end{subfigure}
  \hfill
  \begin{subfigure}[b]{0.49\textwidth}
  \centering
  \includegraphics[width=.8\textwidth]{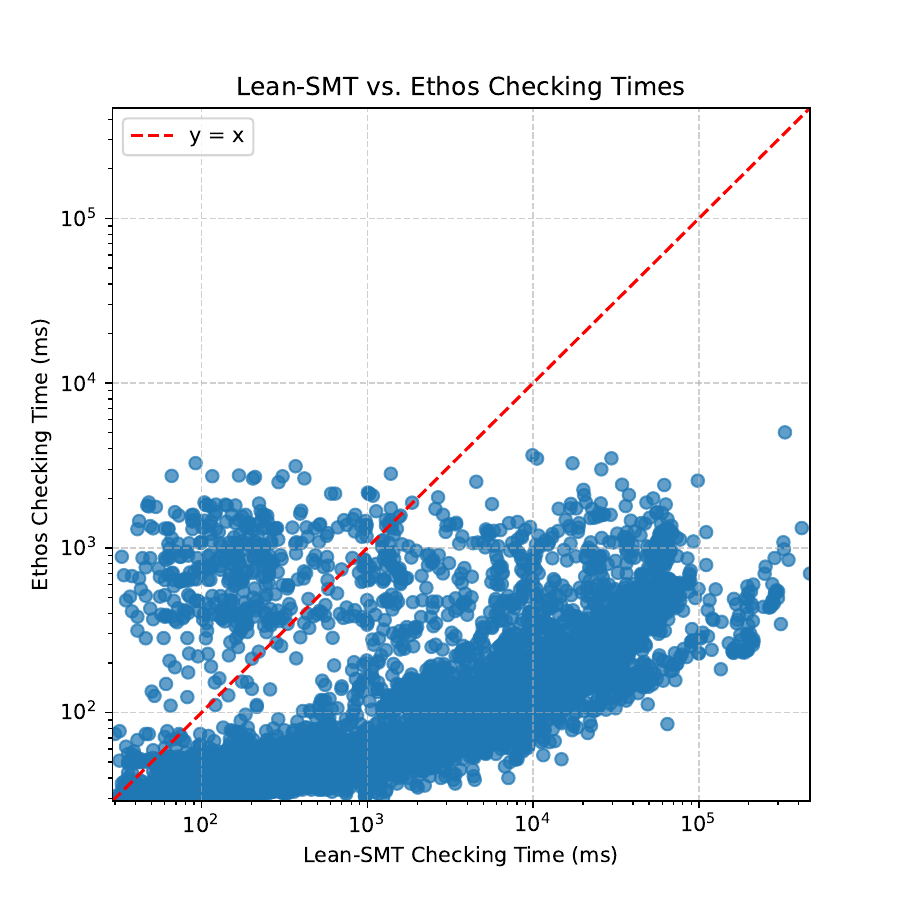}
  \caption{}
  \label{fig:scatter}
  \end{subfigure}
  \caption{(a) shows the performance of \tactic on Sledgehammer benchmarks, while (b) compares proof checking performance of \tactic with Ethos.}
\end{figure}

The \tactic tactic is mainly designed to prove Lean goals provided by users, but can also act as a proof checker for CPC proofs in supported fragments.
We evaluate it\footnote{All benchmarks were executed on a cluster with nodes
  equipped with 48 AMD Ryzen 9 7950X processors running at 4.50GHz and 128GB of
  RAM each.} in these scenarios with the following benchmark sets:
a set of 5000 SMT-LIB benchmarks generated by Sledgehammer from Isabelle/HOL,
taken from \emph{Seventeen Provers Under the
  Hammer}~\cite{DBLP:conf/itp/DesharnaisVBW22}; and $24,817$ unsatisfiable
SMT-LIB benchmarks used in SMT-COMP
2024\footnote{\url{https://smt-comp.github.io/2024/}} that fit the supported
fragments of \tactic: UF, IDL, RDL, LIA, LRA, LIRA, and their quantifier-free
siblings.
The Sledgehammer benchmarks allow us to assess \tactic's performance in both
mathematics and formal verification domains and
include lemmas selected by Sledgehammer with its premise
selection mechanism.
The problems in SMT-COMP are used together with proof-producing SMT solvers to
generate proofs that are passed on to a set of proof checkers, including
\tactic.

\subsection{Isabelle Sledgehammer Benchmarks}

We evaluate \tactic on SMT problems generated
by Isabelle's Sledgehammer.
%
We chose these benchmarks over other options (e.g., Lean's MathLib) due to
Lean's lack of a premise selection mechanism, which is crucial for reducing
false positives (i.e., returning SAT due to missing premises). These benchmarks
also stress test solvers, as they contain many (up to 512) lemmas.
We compare the performance of \tactic against Sledgehammer with the veriT
backend, which supports similar proof reconstruction techniques, and Duper. We do not include CVC4 as a backend for Sledgehammer because its proof production is unstable and do not provide sufficient detail for reliable reconstruction.
The results in Figure~\ref{fig:sledgehammer} show that \tactic
effectively solves a large variety of Sledgehammer benchmarks, underscoring
its potential for integration into general-purpose
proof environments. \tactic outperforms veriT+Sledgehammer mainly because \cvcv
outperforms veriT on this set of benchmarks. \tactic takes less than a second to replay proofs for 98\% of the benchmarks, with the remaining 2\% taking less than 5 seconds. Sledgehammer, by comparison, is faster at reconstructing shorter proofs, but does not scale as well for larger proofs.

\subsection{SMT-LIB Benchmarks}

\begin{figure}[t]
  \centering
  \begin{subfigure}[b]{0.49\textwidth}
    \includegraphics[width=\textwidth]{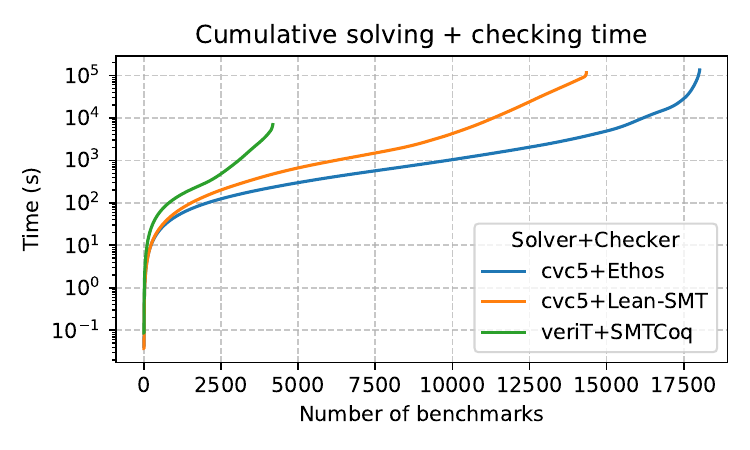}
    \caption{}
    \label{fig:smtliba}
  \end{subfigure}
  \hfill
  \begin{subfigure}[b]{0.49\textwidth}
    \includegraphics[width=\textwidth]{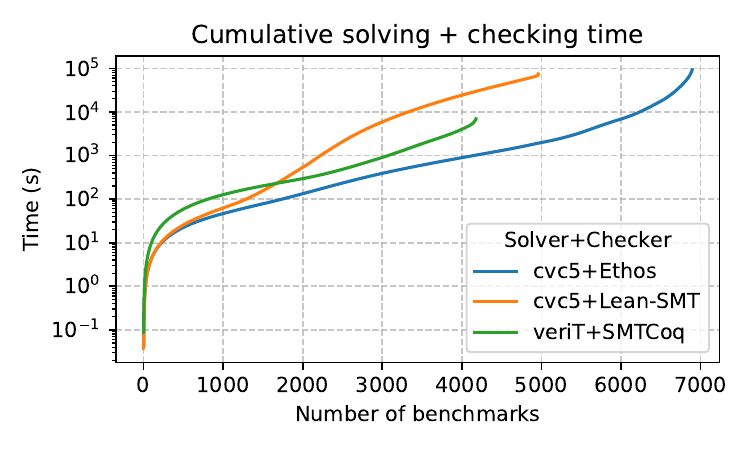}
    \caption{}
    \label{fig:smtlibb}
  \end{subfigure}
  \caption{Figure (a) shows the performance of \tactic on supported SMT-LIB fragments, while Figure (b) shows the performance on the quantifier-free subset.}
  \label{fig:smtlib}
\end{figure}

We evaluate \tactic against the proof checkers
Ethos\footnote{\url{https://github.com/cvc5/ethos/}} v0.1.0 and SMTCoq v2.2.
Ethos is a proof checker implemented in C++ for the Eunoia logical framework, in
which CPC has been formalized.
SMTCoq is a proof checker in OCaml extracted from a formally verified Coq
program, and supports proofs in a subset of the Alethe proof
format~\cite{Schurr2021-Alethe}. It can check
proofs produced by versions of the veriT SMT solver up to 2016.
Since the different approaches use different SMT solvers, our evaluation
includes both proof-checking and SMT solving times. All benchmarks were run with
a standard 20-minute timeout.

Note that both \tactic and SMTCoq are highly trustworthy, since they
both rely on small kernels.
However, SMTCoq's
code extraction mechanism, which extracts OCaml code from verified Coq
code, has to be trusted as well.
The trusted base for Ethos, besides its kernel, also includes the Eunoia
signature for CPC. Moreover, its kernel includes native support for arithmetic
(via GMP) and arrays for efficiency, while Lean's kernel only natively supports
arithmetic.

\paragraph{Supported SMT-LIB Benchmarks}
Figure~\ref{fig:smtliba} compares the cumulative solving and checking times for all SMT-LIB benchmarks.
Out of the $21,595$\footnote{This number is for cvc5+\tactic; for
  cvc5+Ethos, $21,541$ proofs are produced. The difference is due to the
  overhead of proof printing and piping for the latter combo, while in the
  former, the proof is passed directly via the API to \tactic.}
benchmarks proofs generated\footnote{On occasion, cvc5 will produce proofs
  containing holes. Both Ethos and \tactic verify every non-hole proof
  step.} by \cvcv, \tactic successfully verified $15,271$ proofs (71\%),
despite relying solely on the Lean kernel for
soundness. Ethos verifies 98\% of the proofs.
Figure~\ref{fig:scatter} compares the performance of \tactic to Ethos on proof
checking times.
\tactic stays within an order of magnitude of Ethos for most benchmarks. One
reason for Ethos' superior performance is the lack of specialized
support for arrays in Lean's kernel.
This difference could be mitigated by switching to a more efficient array
representation in Lean.

\paragraph{Quantifier-Free Fragment}
Figure~\ref{fig:smtlibb} focuses on the quantifier-free subset of SMT-LIB benchmarks, where SMTCoq's verified approach shines in terms of speed.
However, SMTCoq falls short in the total number of proofs verified, primarily because its rigid, fully verified architecture struggles to keep pace with the rapidly evolving features of modern SMT solvers.
In contrast, \tactic benefits from proof replay flexibility, allowing it to adapt more easily to updates in \cvcv's proof production capabilities, resulting in broader coverage.

\smallskip

Overall, \tactic balances flexibility and performance, achieving promising
results for a proof checker deeply integrated into Lean.
While it trails Ethos in raw speed, its ability to verify a wide range of
benchmarks with a small trusted base makes it an attractive option for checking
SMT proofs in critical domains.

\section{Conclusion}
\tactic, an integration between cvc5 and Lean, is a significant step toward building a Lean hammer that enhances automation and verifies proofs generated by cvc5.
\tactic performs competitively with state-of-the-art proof checkers such as Ethos and Carcara, both in terms of performance and effectiveness.
It is also capable of verifying a diverse range of SMT-LIB benchmarks.
Future work includes creating a dedicated Lean benchmark set for more targeted evaluation and expanding \tactic's support to additional SMT-LIB theories, such as bit-vectors, floats, and strings.
We also plan to extend its proof coverage beyond the current 200 rules and incorporate more preprocessing steps to boost performance. Finally, we plan to improve integration with lean-auto to support higher-order logic and extend support for common Lean datatypes (e.g., tuples, structures, and modular arithmetic).
The ultimate objective is to develop a Lean hammer that brings unprecedented automation and verification capabilities to Lean.

\begin{footnotesize}
  \paragraph{Acknowledgments.}
  This work was partially supported by a gift from Amazon Web Services, the
  Stanford Center for Automated Reasoning, the Coordenação de Aperfeiçoamento de
  Pessoal de Nível Superior - Brasil (CAPES) - Finance Code 001, and the Defense
  Advanced Research Projects Agency (DARPA) under contract FA8750-24-2-1001. Any
  opinions, findings, and conclusions or recommendations expressed here are
  those of the authors and do not necessarily reflect the views of DARPA.
\end{footnotesize}

\bibliographystyle{splncs04}
\bibliography{references}

\newpage
\appendix

\section{Full tactic for reconstructing \texttt{ARITH\_SUM\_UB}}
\label{sec:rec-tactic}

\begin{figure}[t]
\caption{Implementation of the \texttt{sumBounds} tactic}\label{fig:sumBoundsTac}
\begin{lstlisting}[numbers=left]
  def combineBounds (pf₁ pf₂ : Expr) : MetaM Expr := do
    let t₁ ← inferType pf₁
    let t₂ ← inferType pf₂
    let rel₁ ← getRel t₁
    let rel₂ ← getRel t₂
    let opType₁ ← getOperandType t₁
    let opType₂ ← getOperandType t₂
    let (pf₁', pf₂') ← castHypothesis pf₁ pf₂ rel₁ rel₂ opType₁ opType₂
    let thmName : Name :=
      match rel₁, rel₂ with
      | `LT.lt , `LT.lt => `sumBoundsThm₁
      | `LT.lt , `LE.le => `sumBoundsThm₂
      -- Note: omitting other cases
      | _      , _      => panic! "[sumBounds]: invalid relation"
    mkAppM thmName #[pf₂', pf₁']

  def sumBoundsCore (acc : Expr) (pfs : List Expr) : MetaM Expr :=
    match pfs with
    | [] => return acc
    | pf' :: pfs' => do
      let acc' ← combineBounds acc pf'
      sumBoundsCore acc' pfs'
\end{lstlisting}
\end{figure}

With the theorems shown in Section~\ref{sec:proofRec} proven, the tactic shown
in Figure~\ref{fig:sumBoundsTac} can be implemented.
The function \texttt{sumBoundsCore} is meant to be invoked with the last proof in the input
as \texttt{acc} and the rest of them, in reverse order, as \texttt{pfs}.
If \texttt{pfs} is empty, it returns the accumulator (line 18).
Otherwise, function \texttt{combineBounds} is invoked to sum the inequalities
represented by \texttt{acc} and \texttt{pf'} (line 20) and recursively call
\texttt{sumBoundsCore}, updating the arguments (line 21).
The first action performed in function \texttt{combineBounds} is to obtain the
inequalities corresponding to \texttt{pf₁} and \texttt{pf₂} through
\texttt{inferType}. Then, in lines 4 to 7, it inspects their structure and
obtains their relation symbol (\texttt{rel₁} and \texttt{rel₂}) and the type of
their operands (\texttt{opType₁} and \texttt{opType₂}). Notice that
\texttt{sumBoundsThm} expects that all the four variables have the same type. If
one of the input proofs represents an inequality over integers and the other is
over reals, the inequality has to be lifted from an inequality between integers
into an inequality between reals.
This is done using one of the following theorems, depending on
the relation symbol:

\begin{lstlisting}
  Int.castEQ : ∀ {a b : Int}, a = b → (↑a : Real) = (↑b : Real)
  Int.castLE : ∀ {a b : Int}, a <= b → (↑a : Real) <= (↑b : Real)
  Int.castLT : ∀ {a b : Int}, a < b → (↑a : Real) < (↑b : Real)
\end{lstlisting}

The function \texttt{castHypothesis} (line 8) does this analysis and, if necessary,
applies one of these theorems to \texttt{pf₁} or \texttt{pf₂}.
Once with the correct version of the proofs, the respective instance of
\texttt{sumBoundsThm}, depending on \texttt{rel₁} and \texttt{rel₂}, can be
applied.
The \texttt{match} statement in lines 10 to 13 chooses which version to apply.

\section{Tables for for Evaluations}

\begin{table}[t]
  \centering
  \begin{tabular}{lrrr}
    \toprule
    Solver+Checker     & Solved\&Proved & Checked & Checked (no holes) \\
    \midrule
    cvc5+Lean-SMT      & 2868           & 2866    & 2847               \\
    veriT+Sledgehammer & 2211           & 2180    & 2180               \\
    Duper              & 1116           & 1116    & 1116               \\
    \bottomrule
\end{tabular}

  \caption{Performance of \tactic on baseline Seventeen Provers benchmarks (5000 total).}
  \label{tab:sledgehammer}
\end{table}

Table~\ref{tab:sledgehammer} shows the performance of \tactic on the Seventeen Provers benchmarks, while Tables~\ref{tab:smtliba} and \ref{tab:smtlibb} show its performance on SMT-LIB benchmarks.

\begin{table}[h]
  \centering
  \begin{tabular}{lrrr}
    \toprule
    Solver+Checker & Solved\&Proved & Checked & Checked (no holes) \\
    \midrule
    cvc5+Lean-SMT  & 21595          & 15271   & 14099              \\
    cvc5+Ethos     & 21541          & 21196   & 18018              \\
    veriT+SMTCoq   & 4178           & 4178    & 4178               \\
    \bottomrule
\end{tabular}

  \caption{Performance of \tactic on supported SMT-LIB fragments (24817 total).}
  \label{tab:smtliba}
\end{table}

The first column in the tables shows the number of benchmarks solved by solved by the ATP. The time it takes to generate the proof is also included. The second column show the number of benchmarks where proof checking was successful. This includes proofs with holes, which are not considered by SMTCoq. The third column shows the number of benchmarks where proof checking was successful and the proof was complete.

\begin{table}[t]
  \centering
  \begin{tabular}{lrrr}
    \toprule
    Solver+Checker & Solved\&Proved & Checked & Checked (no holes) \\
    \midrule
    cvc5+Lean-SMT  & 9263           & 5091    & 4869               \\
    cvc5+Ethos     & 9203           & 8860    & 6892               \\
    veriT+SMTCoq   & 4178           & 4178    & 4178               \\
    \bottomrule
\end{tabular}

  \caption{Performance of \tactic on the quantifier-free subset of SMT-LIB benchmarks (11804 total).}
  \label{tab:smtlibb}
\end{table}

\end{document}